# Comments to a series of works by V.K. Mukhomorov on the theory of a continuum polaron and two-center bipolaron (axially symmetrical quasimolecular dimer)

N. I. Kashirina[*], V. D. Lakhno[**]

*Institute of Semiconductor Physics of the National Academy of Sciences of Ukraine
41, prospect Nauky, Kiev 03028, Ukraine*

**Institute of Mathematical problems of Biology, Russian Academy of Sciences,
Pushchino, Moscow region, 142290, Russia*

***Abstract.*** Some critical remarks are made regarding a series of works by V. K. Mukhomorov dealing with polarons and oscillatory and rotational spectrum of a large-radius bipolaron near a subsidiary minimum corresponding to its two-center configuration. It is shown that V. K. Mukhomorov's conclusion that by varying the bipolaron functional one should look for a constrained rather than absolute minimum is erroneous. Consideration of interelectronic correlations corresponding to a direct dependence of the wave function of the studied system on the distance between the electrons does not break the virial theorem. In the strong coupling limit the virial theorem holds true for both one-center and two-center bipolaron states, the latter representing a subsidiary minimum which arises due to a choice of an insufficiently versatile variation function. The work is done with the support from the RFBR (project № 07-07-00313).

## INTRODUCTION

The theory of polarons and bipolarons (BP) which was initially developed to explain some physical phenomena in polar dielectrics plays an important role in many physical, chemical and biological processes occurring not only in solid bodies but also in organic compounds and polar liquids including water and various water solutions. In [1- 7] a polaron theory of charge transfer in proteins was developed and in [8] this theory was used to describe a hydrated electron.

In works [9, 10] by A. S. Davydov, the fundamentals of the translation-invariant theory of a continuum polaron developed by N. N. Bogolubov [11] and S. V. Tyablikov [12] were used as the basis for the theory of solitons in biological molecules. At the present time this application of the polaron theory is being studied intensively [13, 14, 15]. In [16-22] a dynamical theory of soliton formation and soliton transfer in DNA was developed.

In the continuum approximation, within the framework of which the electron-phonon interaction is described in the theory of polarons and large-radius bipolarons, use is made of Fröhlich Hamiltonian, and the electron effective mass serves to represent the electron mass. This approach is valid for both ionic crystals and amorphous media or polar liquids. The latter is possible by virtue of the fact that the structure of the medium, in which quasifree electrons and their large-radius autolocalized formations move, does not affect the character of the process concerned. Indeed, the continuum approximation only uses the macroscopic parameters of a medium such as the effective mass of charge carriers or high-frequency and static dielectric permittivities, while electric fields induced by long-wave optical phonons smoothly vary at the wavelengths comparable with distances between ions in polar media.

---

[*] n_kashirina@mail.ru
[**] lak@impb.psn.ru

Unfortunately, the vast majority of Russian journals lack a special section devoted to criticism of erroneous trends which are developed in many papers and published in various special editions. Here we will try to fill this vacuum in the theory of polarons and large-radius bipolarons and make some critical remarks regarding a series of works by V. K. Mukhomorov concerned with polaron subject-matter. V. K. Mukhomorov's papers on a two-center bipolaron other than those discussed in this paper can be found in the references in these works.

Although the polaron theory dates back more than half a century, and numerous mistakes that were inevitably made in the course of development of modern scientific concepts were corrected, at the present time a trend of the bipolaron theory which seems to be devoid of physical meaning, namely, the continuum theory of a two-center bipolaron, is still being intensively developed. This is especially true in regard to numerous works dealing with oscillatory and rotational spectra of a continuum two-center bipolaron, or an axially symmetrical quasimolecular dimer, as this system is termed by V. K. Mukhomorov. Recently, some works co-authored with V. K. Mukhomorov have been published [23, 24] in which erroneous ideas concerned with a two-center bipolaron were used as the basis for constructing a pair potential of interaction between polarons. The latter, notwithstanding a correct general formulation of the problem, leads to qualitative and quantitative mistakes in describing the physical phenomena involved.

The authors of [25-30] showed that at the parameters of the medium at which a bipolaron can be formed, the minimum corresponding to the two-center configuration represents a shallow subsidiary minimum which disappears when the wave function (WF) of the studied system is chosen more adequately. Consideration of interelectronic correlations associated with a direct dependence of the WF of the system on the distance between the electrons leads to a qualitative change in the form of the inter-polaron interaction potential. In this case the shallow minimum corresponding to a bipolaron two-center configuration disappears and the power dependence has only one minimum corresponding to a one-center bipolaron, or Pekar bipolaron.

To our knowledge, the first work in which a one-center bipolaron was successfully studied with due consideration of interelectronic correlations was that by Suprun and Moizhes [31]. Earlier Larsen [32] showed that taking into account of interelectronic correlations associated with a direct dependence of the WF on the distance between the electrons leads to a huge increase in the coupling energy of $D^-$ – center or a bipolaron coupled on Coulomb potential analogous to $H^-$ - ion in atomic physics.

The reasons for which a two-center large-radius bipolaron still remains the subject of V. K. Mukhomorov's studies are detailed in [33] where, apart from the problems concerned with erroneous investigations of a bipolaron spatial configuration, as will be shown in what follows, he formulated some propositions inconsistent with the fundamentals of both the polaron theory and quantum mechanics. We mean erroneous treatment of the variational principle and virial theorem as applied to calculations of the bipolaron energy. In this connection we focus our attention on criticism of the erroneous statements made in that paper. Besides, we make some critical remarks regarding some other works of the author on the theory of polarons and bipolarons.

In [30] we thoroughly analyze the conditions at which the virial theorem holds true for a strong coupling bipolaron. There we deal with both one-center and two-center configurations as well as with the most general case, i.e. the two-center model, with due regard for interelectronic correlations.

# VARIATIONAL PRINCIPLE AND VIRIAL THEOREM IN THE POLARON THEORY

In [33] V. K. Mukhomorov states that the calculations of the bipolaron energy made in [25, 31, 34] (from here on the references are cited in accordance with the numeration of this work) with due regard for a direct dependence of the WF of the studied system on the distance between the electrons yielded lowered values of the ground state energy, and as a consequence, overestimated values of the bipolaron coupling energy. This opinion is reasoned by the fact that the authors of [25, 31, 34] found an absolute minimum of the functional corresponding to a strong-coupling bipolaron, rather than a constrained one, as they should have done according to the rule suggested by V. K. Mukhomorov.

In other words, V. K. Mukhomorov argues that the use of a direct variational method for finding the minimum of a functional corresponding to the energy of a quantum mechanical system can lead to lowered values of the ground state energy. This statement is in conflict with a fundamental of the quantum mechanics which holds that "in the ground state, the energy of a system is the lowest minimal value, i.e. the absolute minimum" (see [35], p. 156), that is the energy of a system $E$ is determined as the lower boundary of $E = \langle \Psi | H | \Psi \rangle \langle \Psi | \Psi \rangle^{-1}$, where H is a Hamiltonian of the system.

Let us replace the probe function $\Psi_0(\mathbf{r})$ (where $\mathbf{r}$ stands for $\mathbf{r_1\ r_2,...r_N}$, N is the number of electrons) by the function $\Psi = \lambda^{3N/2} \Psi_0(\lambda \mathbf{r})$. The scaling coefficient $\lambda$ will be considered as a variational parameter. In so doing we do not impose any additional limitations on the initial functional, but, on the contrary, introduce the additional variational parameter $\lambda$ and seek an absolute (rather than constrained, as is suggested in [33]) minimum of functional (1) with the WF $\Psi$. Obviously, relations $E \leq J[\Psi_0]$; $E \leq J[\Psi]$ are fulfilled, where $E$ is the ground state energy. In the general case we cannot make a categorical conclusion of which of the two values - $\min J[\Psi]$ or $\min J[\Psi_0]$ - is larger, since during numeral minimization it is possible to appear in distinct local minima, corresponding to these functionals. Therefore the use of a scaling transformation does not restrict from below the energy of a functional being minimized. When the function $\Psi_0(\mathbf{r})$ leads to an extremum of the functional, at $\lambda = 1$ $\Psi$ turns to $\Psi_0$, and $J[\Psi]$ must have an extremum at $\lambda = 1$ [36, 37]. This property of a scaling transformation is so general that holds for any extremum of the initial functional $J[\Psi_0]$: maximum, minimum, or any local extremum. We intentionally refer to a paper by S. I. Pekar and M. F. Deigen [37] who clarify the essence of the scaling transformation used in classical works on the interaction of electrons with a phonon field. The use of the properties of the initial Hamiltonian enables one to carry out variation in terms of the additional variation parameter $\lambda$ "by hand" and yields the well-known relation (valid for both a polaron and a bipolaron) which implies that the total ground state energy corresponds to the minimum of the functional $E = -\min((V_q + V_f)^2/4T)$, where $E$ is the ground state energy, $T$ is the kinetic energy, $V_q$ is the energy of an interelectronic interaction (for a polaron $V_q = 0$), $V_f$ - is the energy of electron- phonon interaction.

V. K. Mukhomorov's statement that the criterion of optimality of electron probe functions is the fulfillment of the virial theorem is erroneous since the theorem in itself is fulfilled for any extremum of the studied functional including maxima and all the local minima. Naturally, in S. I. Pekar's works there is nothing like what V. K. Mukhomorov says referring to the monograph [38]. On the contrary, S. I. Pekar's statement is as follows: "As is known from the theory of direct variational methods, one should assign meaning only to the lowest extremum value of a studied functional… All the other extrema can turn out to be a result of insufficient versatility of the approximation function and disappear when passing on to more general approximations" [39, p.67] (monograph [38], which has become a bibliographic rarity and has been included almost completely in S. I. Pekar's selection [39]).

All the manuals to which V. K. Mukhomorov refers in [33], including classical works of the 50-s and still earlier papers provide an explanation to the fact that the scaling transformation procedure concerned with changing to a new probe function by the replacement $\Psi(\{\mathbf{r}\}) \to k^{3N/2}\Psi(\{k\mathbf{r}\})$ can lead to a "considerable decrease of the energy" [36, p. 223], and not the reverse, as is stated in [33]. Therefore, if it had turned out that in our calculations [25] or in calculations of other authors [31] cited in [33] as erroneous, the virial theorem does not hold, this would have indicated that the minimum of the functional is not found and the correct energy value should be lower, and not the reverse.

Moreover, the virial theorem holds true even if we would vary an erroneous functional in which some terms are missing. As a rule, in this case the values of the ground state energy are considerably lowered. If a work with an erroneous result is published, the calculations are reproduced by other authors and the errors are eliminated.

In [40], where the energy of a one-center bipolaron was calculated, some terms were missing in the kinetic energy. This led to a considerable decrease in the ground state energy as compared to other calculations. In particular, this value was much less than that in [31]. The calculations were reproduced by several authors [41 – 43] and the error was corrected.

Later on the method for calculation of a bipolaron energy suggested in [40] was successfully developed in [44] (which was concerned with a one-center bipolaron) and yielded one of the lowest values of the total ground state energy of a bipolaron. V. K. Mukhomorov erroneously cites [44] as a work on a two-center bipolaron performed by alternative methods.

## PEKAR BIPOLARON AND ELECTRONIC CORRELATIONS
## (HISTORY AND FORMULATION OF THE PROBLEM)

The variational functions used in [31] were earlier [45] applied to calculate the $F'$- center energy. S. I. Pekar offered his post-graduate student O. F. Tomasevich to calculate the $F'$-center energy taking into account correlations, related to the direct dependence of WF on a distance between electrons.

S. I. Pekar never carried out such calculations himself, and his conclusion that taking into account of interetectronic correlations reduces the calculated value of the bipolaron energy by not more than 1-2% [38, 39, p. 124] was made with reference to O. F. Tomasevich's calculations published in [45].

The same function was used in [46] to calculate the bipolaron energy in a metal-ammonia solution (we give a correct reference to A. S. Davydov's paper, since in [33] reference [3] to this paper is erroneous). It is interesting, that A. S. Davydov did not perform such calculations either. As is reported in [46], (p. 7) they were made by a Kiev State University student, Rozenblat.

A short message by S. G. Suprun and B. Ya. Moizhes about stability of Pekar bipolaron (just this model was suggested in monograph [38]) kindled the interest of Kievan physicists who were acquainted with the beginnings of the bipolaron theory. V. I. Vinetsky offered one of the authors (N. I. Kashirina) to reproduce O. F. Tomasevich's calculations. They were reproduced and revealed that in [45] the normalizing integral N (which is the simplest one) was calculated erroneously. Since the phonon part of the functional $V_{Bp(f)} \sim N^{-2}$, just this negative term providing the stability of the $F'$-center and the bipolaron coupling energy was considerably underestimated in [45]. Unfortunately this error belonged among the rare cases when the coupling energy was underestimated rather than overestimated, as in [40], at the same time, passage to the limit of the functional without taking into account interelectronic correlations was fulfilled correctly.

The results of [45] were only corrected in [31]. The title of the paper "On the role of electronic correlation in the formation of Pekar bipolaron" [31] indicates that its authors recognize S. I. Pekar as the creator of the model of a one-center bipolaron, the existence of which they have managed to prove in more that thirty years after S. I. Pekar had suggested the model, including the WF used for the variational calculations in [31].

Not long before the paper on a one-center bipolaron [31] came out, Larsen [32] calculated the $D^-$ center energy by Buymistrov-Pekar method [47] (BPM). There he reported that consideration of interelectronic correlations leads to a giant increase in the $D^-$ center coupling energy. The WF`s used in his work were very close to those applied by S. G. Suprun and B. Ya. Moizhes [31] for calculating the bipolaron energy.

Therefore, in the pioneering papers on a one-center bipolaron [45, 46], a conclusion about a lack of any coupled states was drawn as a result of a trivial numerical error made in calculations of the bipolaron ground state functional, but not in the least owing to the fact that there, as distinct from [25, 31], the virial theorem was used, as is stated by the author of [33].

In paper [48] devoted to calculations of the one-center bipolaron energy, it is reported that in finding the minimum of the functional use is made of a scaling transformation at which the virial theorem holds true automatically. In other papers this may not be reported, as it is not always reported what particular method was used to find the minimum of the multiparameter functional.

## TOTAL ENERGY OF INTERELECTRONIC INTERACTION AND BIPOLARON SPATIAL CONFIGURATION

The minimum obtained in the framework of a one-center model, could prove to be a maximum on the curve for the energy dependence of two polarons if we introduce an additional variational parameter which describes the axial symmetry of the bipolaron WF and plays the role of the distance between the polarization wells of two interacting polarons. In this case we could get a lower value of the bipolaron energy. With this aim in view we carried out variational calculations with a maximally versatile WF given in [25]. To describe qualitatively the disappearance of subsidiary extrema associated with gradual decrease of the WF versatility, we can restrict ourselves to only one term in the WF (4) given in [25]. These results do not require invoking any complicated programs for calculating the minimum of the multiparameter functional, neither do they demand high-power computers and can be easily reproduced. In fig. 1 we present some curves calculated for the simplest WF

$$\Psi_{12} = N(1 \pm P_{12})\exp(-a_1 r_{a1}^2 - 2a_2 \mathbf{r}_1 \mathbf{r}_2 - a_3 r_{b2}^2), \quad (1)$$

where $N$ is a normalizing multiplier, $P_{12}-$ is an operator for transposition of electron coordinates, $a_1, a_2, a_3$ are variational parameters used for finding the energy of a two-center bipolaron as a function of a distance between the centers of polarization wells of two polarons.

Curve (1) corresponds to absolute absence of interelectronic correlations associated with a direct dependence of the WF on the distance between the electrons ($a_2 = 0, a_1 = a_3$) and at the point $R = 0$, where $R$ is a distance between polarons, changes to the value corresponding to doubled polaron energy calculated with one Gaussian function. Curve (2) corresponds to a more versatile WF, when a direct dependence on the distance between the electrons is lacking, but in view of the fact that $a_1 \neq a_3$ at $R = 0$ the WF does not change to a product of one-electron WFs, i.e. retains the unmultiplicative property. It is seen that in this case, at $R = 0$ a nonzero energy of a bipolaron coupling appears which corresponds to a minimum shallower than that at $R \neq 0$. For curve (3) it was assumed that $a_2 \neq 0$, $a_1 = a_3$. Subsidiary minimum corresponding to a two-center bipolaron disappears; the minimum at the point $R = 0$ deepens. And finally, for the case $a_2 \neq 0$, $a_1 \neq a_3$, which is represented by curve (4) we have a deeper minimum at $R = 0$.

It should be noted, that the virial theorem holds true for the four curves in fig. 1, that is for all R relation (5) given in [33] is valid, i.e.

$$R \frac{dE(R)}{dR} + 2T(R) + U(R) = 0,$$

where $E(R)$, $T(R)$, $U(R)$ - are total, kinetic and potential energies of the bipolaron correspondingly, $R$ - is the distance between the centers of polaron polarization wells.

The total dependencies of the bipolaron energy on the distance between the centers of polarization wells of two polarons (curves (3) and (4) in fig.1) obtained by us for the parameters of crystals in which the criterion of the existence of a bipolaron coupled state is met, have a single minimum at $R = 0$, which corresponds to a spherically symmetrical formation, rather than to an axially-symmetrical one. The latter is in complete agreement with the qualitative inference and the dependence of the total energy of the interaction between two polarons presented in a review by Mott [49, fig.2] and does not contradict any general principles of the multielectron theory, as is stated in [33].

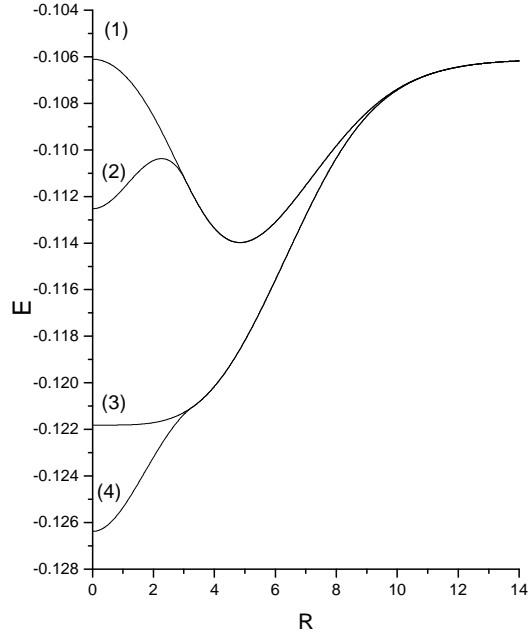

**Fig. 1.** Dependencies of the bipolaron total energy on the distance between the centers of polarization wells of two polarons for the probe WF (1) in the strongest coupling limit $\eta = 0$. Curves (1) ÷ (4) correspond to the following parameters (1) – $a_2 = 0, a_1 = a_3$; (2) – $a_2 = 0, a_1 \neq a_3$; (3) – $a_2 \neq 0$, $a_1 = a_3$; (4) – $a_2 \neq 0$, $a_1 \neq a_3$. The unit of energy is $Ha^* = \hbar^2 / m^* a_0^2$, the unit of length is $a_o = \hbar^2 \varepsilon_\infty / m^* e^2$.

Similar dependencies of the energies of $H_2^+$ and $H_2$ molecules on the distance between the protons obtained without regard for a repulsion between the nuclei [35] (fig. 3, p. 30), [50] (fig. 3.3., p. 76) are well known. The latter enables one to study the energy spectrum of a helium atom or a hydrogen molecule within the framework of the same model.

A minimum appears on the curve for the dependence of the hydrogen molecule energy on the distance between the protons [51] (fig. 56, p. 362) only when a repulsion between the protons is taken into account. Therefore, taken alone, the dependence of the WF on the distance between the protons does not lead to occurrence of a minimum at the point $R \neq 0$.

In the Hamiltonian of a system consisting of two polarons, the term representing a repulsion between the centers of polarization wells is lacking. Moreover, even the Fröhlich Hamiltonian for a two-electron system is independent of the distance $R$ between the centers of polarization wells. This dependence appears only when the probe WF is chosen to contain an axial component. Therefore no wonder that an extremely small minimum at $R \neq 0$ just disappeared when a more versatile WF was chosen.

## BIPOLARON EXCITED STATES

The possibility of an existence of excited, or two-quantum states of bipolarons which can appear when we consider the energy of an interaction between two polarons, one of which is in an excited state (two-quantum 2s or 2p states), and the other is in the ground $1s$-state, deserves a special study which will be carried out in an independent work. Notice, however, that in [52] the energy corresponding to a relaxed two-quantum bipolaron state in which one electron is in the ground 1s- state and the other - in the excited 2p – state (in what follows we put $\eta=\varepsilon_\infty/\varepsilon_0=0$) $^3F = -0.08943 Ha^*$ is greatly lowered.

We reproduced calculations of the bipolaron energy corresponding to this state for the functions determined by expressions (3), (4) in work [52]. For one-electron WFs 1s and 2p chosen in the form of $S(1) = (1 + \alpha \cdot r_1 + \gamma \cdot r_1^2) \exp(-s \cdot r_1)$ and $P(r) = (1 + dr) z \exp(-pr)$ we obtained $^3F = -0.0758146 Ha^*$. At $\gamma = 0$ in WF S(1) we got $^3F = -0.0750143 Ha^*$. The latter result corresponds to WF (4) in [52] and, as we might expect, exceeds our value of the bipolaron energy in the lowest triplet state $^3F = -0.076072 Ha^*$ which was obtained with the use of 5 Gaussian orbitals [25]. At the same time the energy of a polaron in 2p – state $F_p = 0.0197 Ha^*$ is overestimated in [52]. Our result obtained with the use of a WF P(r) coinciding with the WF (4) from [52] was $E_p = -0.021002 Ha^*$. For comparison we present the value of the energy of a relaxed p-state obtained in [53] as a result of numerical solution of a relevant Euler equation $E_p = -0.02285 Ha^*$.

Both the errors lead to a considerable increase of the bipolaron coupling energy, especially with regard for the fact that, according to the rule suggested by V. K. Mukhomorov in [52], the bipolaron coupling energy in an orthostate is determined by the difference $F_s + F_p - {}^3F$ (where $F_s$, $F_p$ are the polaron energies in the ground state and 2p –state, respectively), rather than by $2F_s - {}^3F$, as in our work [25]. In this connection we believe that conclusions about the possibility of the existence of a metastable one-center bipolaron in a triplet state are made in [52] as a result of numerical errors.

## TWO-CENTER BIPOLARON IN A STRONG COUPLING METHOD

To make sure that in the strong coupling limit V. K. Mukhomorov deals with the same functional as the authors of [25, 31, 34] we compared functional (1) from [33] (which, according to V. K. Mukhomorov was obtained in [55] using the results of the translation-invariant theory of a strong-coupling bipolaron)

$$E(R) = -\frac{\hbar^2}{2m^*} \int_{\tau_1=\tau_1'} \nabla_1^2 \rho_1(r_1, r_1') d\tau_1 + \frac{1}{4} \iint d\tau_1 d\tau_2 \rho_2(r_1, r_2)$$

$$\times \left\{ 2g(r_1, r_2) \varepsilon_\infty^{-1} + \varepsilon^{*-1} \sum_{i=1,2} \int g(r_i, r_i') \rho(r_i') d\tau_i' \right\}$$

with functional (15) from ref. [54]. This BP functional was obtained by Buymistrov-Pekar method for the case when translation symmetry is lacking. A similar functional is also used by V. K. Mukhomorov in work [56]. The expression $E(R)$ presented above describes a strong coupling addition to the complete functional. As it should follow from [54, eq. (16)], before the summation over the wave vectors this addition is written as

$$J_{Bp}[\varphi] = -\sum_{j=1,2}\frac{\hbar^2}{2m^*}\int|\nabla_j\varphi(\vec{r}_1,\vec{r}_2)|^2 d\tau_1 d\tau_2 + \left\langle\varphi\left|\frac{e^2}{\varepsilon_\infty r_{12}}\right|\varphi\right\rangle$$

$$-\sum_{\vec{f},j=1,2}\frac{2A_f^2}{\hbar\omega_f}\langle\varphi|\exp(-i\vec{f}\,\vec{r}_j)|\varphi\rangle\langle\varphi|\exp(i\vec{f}\,\vec{r}_j)|\varphi\rangle$$

That is $E(R) = J_{Bp}[\varphi]$ and both the functionals coincide with the strong coupling functional used in ref. [25, 31, 57, 58] up to passing on from summation to integration over the phonon wave vectors. The error concerned with incorrect determination of the kinetic energy in eq. (16) in [54] was corrected in [33, eq. (1)]. Therefore if we exclude technical misprints (functional (1) in [33] has a needless multiplier 1/2 which must be absent in determining the density matrix with a normalized Hartry-Fok function [36], p. 215) and pass on to the notation used in [25, 31], then it becomes clear that we deal with a traditional strong coupling functional for two electrons in a phonon field

$$E(R) = T_{12} + J_{12}/\varepsilon_\infty - 2J_{13}/\tilde{\varepsilon}, \qquad (2)$$

where, $T_{12} = -\hbar^2/2m^*\int\Psi_{12}(\Delta_1+\Delta_2)\Psi_{12}d\tau_{12}$; $J_{12} = \int(r_{12}^{-1}\Psi_{12}^2)d\tau_{12}$; $J_{13} = \int(r_{13}^{-1}\Psi_{12}^2\Psi_{34}^2)d\tau_{12}d\tau_{34}$; $\tilde{\varepsilon}^{-1} = (\varepsilon_\infty^{-1} - \varepsilon_0^{-1})$; $\Psi_{ij} \equiv \Psi(r_i, r_j)$, i, j=1,2; $\varepsilon_0$, $\varepsilon_\infty$ are static and high-frequency dielectric permittivities, respectively.

For Heitler-London WF, this expression exactly coincides with the part of the functional from ref. [59] corresponding to a two-electron system in a phonon field, describing an exchange-coupled pair of paramagnetic centers interacting with optical phonons in ionic crystals. Having omitted the terms for the interaction of an electron with Coulomb centers we get a functional of a two-center bipolaron [57, 58], or an "axially-symmetrical quasimolecular dimer" as it is termed by V. K. Mukhomorov in [54, 60]. Clearly, this functional bears no relation to translation-invariant solutions of Fröhlich equation for a bipolaron, let alone the fact that it was obtained not in V. K. Mukhomorov's works, as it is stated in [33].

Numerical calculations carried out in [33] demonstrate clearly that for the WFs dealt with by V. K. Mukhomorov, «the one-center state of a bipolaron is not stable under any circumstances which do not violate the main physical principles" [33]. This statement has no need in any proof in view of the fact that, in the limit of the strongest coupling, when $\eta = \varepsilon_\infty/\varepsilon_0 = 0$, for the multiplicative WF the bipolaron functional splits into a sum of two noninteracting polarons. In the general case, for multiplicative WFs if we add into (2) the identically zero expression $J_{12}/\varepsilon_0 - J_{12}/\varepsilon_0$ and rearrange the terms, the bipolaron energy $E_{Bp}$ will be expressed as

$$2E_p \leq E_{Bp} = 2E_p + J_{ss}/\varepsilon_0, \quad J_{ss} = \int a(1)^2 a(2)^2 r_{12}^{-1} d\tau_{12}, \qquad (3)$$

where $a(1)$ are one-electron orbitals.

This property holds not only for the strong coupling functional, but also for bipolaron functional obtained by the intermediate coupling method. Therefore, in this approximation for $\eta = 0$ at $R \to 0$ the bipolaron energy must tend to the same limit as it does at $R \to \infty$, since the bipolaron WF chosen in the Heitler-London form at the point $R = 0$ changes into a multiplicative product of one-electron functions.

## THE USE OF BUYMISTROV-PEKAR METHOD FOR CALCULATING THE ENERGY OF A TWO-CENTER BIPOLARON

The bipolaron method suggested in [47] for the case when a translation invariance is lacking, was used in [54, 60] to consider a two-center bipolaron. For the most part, the text of these works reproduces that of paper [47] which deals with a two-electron system in an ionic crystal, up to equation (18) in [47]. Since V. K. Mukhomorov passes on from Fröhlich

Hamiltonian and the standard designations to the Hamiltonian used in [47] the derivation of the final expression (18) from [47] takes more space than in the original paper. Formula (18) is written in [47] with regard to the fact that correlation effects are absolutely lacking, and the WFs are chosen in the multiplicative form (one-center system, Pekar bipolaron), therefore the part of the two-electron functional corresponding to the addition of an intermediate coupling is simplified and is written in the form of a sum of relevant additions of two polaron functionals. Nevertheless V. K. Mukhomorov erroneously inserts in this functional a non-multiplicative two-electron function by Heitler-London method. In order to introduce the strong coupling, he uses the general form of the functional which was earlier studied by Vinetsky [58] who dealt with one-electron functions chosen in the form (20) from [54]. The functional for intermediate coupling addition in [54, eq. (14) and (15)] contains an error, since the non-multiplicative functions should have been inserted in quite a different expression (eq. 10 from [47]). For this reason in [54, 60] all the terms containing the form $\langle \Psi_{12} | \exp(-i\mathbf{k}\mathbf{r}_{12}) | \Psi_{12} \rangle$ are lost in the intermediate coupling addition, that is the part of the functional corresponding to the intermediate coupling is determined by the erroneous formula. As for the strong coupling functional from ref. [54, 60], it had already been investigated by V. L. Vinetsky in work [58] and had not led to a considerable widening of the bipolaron existence domain as compared to the calculations in which the simplest hydrogen-like functions were used.

At the point $R=0$ ($\eta=0$) the bipolaron coupling energy must identically turn to zero since there the Heitler-London functions change to a product of one-electron orbitals and the bipolaron functional breaks up into a sum of two polaron functionals. In V. K. Mukhomorov's work, on the contrary, the maxima on the curves for the bipolaron coupling energy $\Delta F = F_{Bp} - 2F_P$ at the point $R=0$ correspond to negative values and fall to still lower energies as the coupling constant decreases, and at $R \to \infty$ start demonstrating correct asymptotic behavior, i.e. vanish (fig. 1 in [60], and coinciding with it fig. 1 in [59]). Therefore in the sited works, the limit transitions are not fulfilled which indicates that they contain numerical errors.

## VIOLATION OF VIRIAL THEOREM IN V.K. MUKHOMOROV'S WORKS

Analysis of fig.1 in paper [33] can serve to illustrate that the virial theorem does not hold true there, rather than prove that a one-center bipolaron is not stable. Indeed, curve (1) corresponding to the bipolaron kinetic energy in Hartry-Fok approximation, at $R=0$ changes to double polaron energy with opposite sign. At the same time, as is seen from fig.1 in [54, 60], at the same point, the energy of the bipolaron ground state in the strong coupling limit is not equal to double bipolaron energy in the same Hartry-Fok approximation. The numerical results in [54] were reported to have been obtained approximately. At first the strong coupling functional was varied, and then the parameters obtained were substituted into the addition for intermediate coupling. This approximation is justified for relatively high values of the parameter $\alpha$. Thus, the virial theorem must hold true for the part of bipolaron functional corresponding to the strong coupling, and according to this theory, at the extremum point at $R=0$ (point of maximum in the case under consideration) the total energy must be exactly equal to kinetic energy with opposite sign.

Dependencies (2) and (3) in fig.1 of [33] describing consideration of correlation terms in the total functional are also erroneous. Fig.1 in [33] suggests that at $R=0$ the kinetic energy decreases as the correlation effects increase. According to the virial theorem this means that the total energy of the system increases when the electronic correlations are taken into account. Let us remark here that this behavior of the kinetic energy cannot be accounted for by a trivial technical misprint in fig.1 in [33], since in the text of [33] V.K. Mukhomorov interprets fig.1 as follows: "It is seen from fig.1 that as the versatility of the electron WF increases, the general dependence, i.e. decrease of the contribution into the electron kinetic energy at $R \to 0$ is preserved". Notice that parameter $C_1$ is variational (see eq.(9) in [33]), and, as the functional varies, it should be put equal to zero since, as it follows from fig.1 in [33], inclusion of the electron configuration $2p$ into the WF leads to a higher energy. Maybe in [33] in this case too, in exact antithesis with the virial theorem, the total energy decreases as the kinetic energy falls down? Indeed, at page 817 [33] one can read that "at zero the total bipolaron energy decreases due to inclusion of the electronic correlation into the total function".

The question arises of whether V.K. Mukhomorov carried out any numerical calculations at all, or the dependencies of the bipolaron coupling energy on the distance between the polarons given in [33, 54, 60] were obtained from qualitative assumptions in which the main properties of the system under consideration were not taken into account?

In fig.2 we show the dependencies of the bipolaron kinetic energy obtained by us for the same parameters of the WF as in fig.1. It is seen that the kinetic energy curves occur in reverse order as compared to the curves representing the total energy of the system and at $R = 0$ are equal to the bipolaron potential energy with opposite sign.

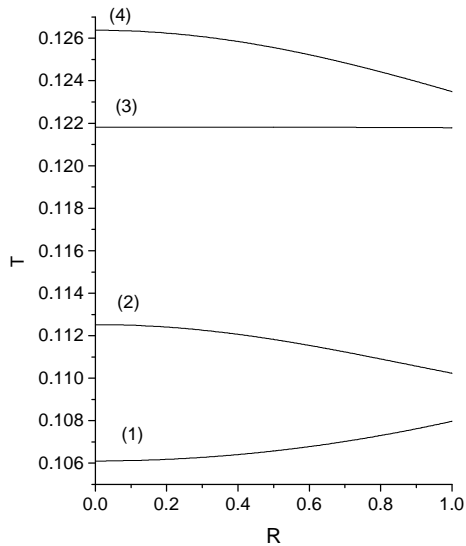

**Fig. 2** Dependence of the bipolaron kinetic energy on the distance between the centers of polarization wells of two electrons for probe WF (1) and parameter $\eta = 0$. The energy dependencies (1) ÷ (4) correspond to the same variation parameters values as in fig. 1.

## POLARON ENERGY IN BUYMISTROV-PEKAR METHOD
## (TRANSLATION SYMMETRY IS LACKING)

The calculations of the polaron energy by Buymistrov-Pekar method presented in [54, 60] for determination of the bipolaron coupling energy should also be recognized as erroneous. In these works V. K. Mukhomorov states that in the intermediate coupling range ($\alpha_c \sim 5 \div 10$) Buymistrov-Pekar method yields lower values of the polaron energy than integration over Feynman trajectories does. But actually Buymistrov-Pekar method leads to energies lower than in Feynman method only in the strongest coupling limit ($\alpha_c \geq 30$) and only due to the fact that this method enables one to choose the WF in a more accurate form and the polaron energy in Feynman method in the limit of large $\alpha$ approaches the result obtained by direct variation of a one-parameter Gaussian function. At $\alpha \leq \alpha_c = 6$, the polaron coupling energy obtained by Buymistrov-Pekar method [47], which V. K. Mukhomorov uses in his works, obeys a simple dependence $J_p \approx -\alpha \hbar \omega$ given in [61]. This value of the polaron energy is much higher than that obtained by Feynman. Paper [34], cited by V. K. Mukhomorov as erroneous, presents a correct dependence of the polaron energy, obtained by Buymistrov-Pekar method, on the constant of electron-phonon interaction. Paper [26], where the energy of an intermediate coupling bipolaron is calculated by Buymistrov-Pekar method [47], presents a comparison of polaron energies obtained by Buymistrov-Pekar method and other techniques.

At the same time, to study the region of a bipolaron existence, V. K. Mukhomorov not only fails to use the value of Feynman energy, but he also takes a value considerably exceeded as compared to that found by Buymistrov-Pekar method. For example, in [54], at $\alpha = 5$, the polaron energy is calculated to be $J_p = -(1.622 + 0.1072\alpha^2)\hbar\omega = -4.3\hbar\omega$ instead of $-5\hbar\omega$, as it must be in Buymistrov-Pekar method for this constant of electron-phonon coupling irrespective of whether Gaussian, or hydrogen-like functions are chosen as probe functions of polarons. If V. K. Mukhomorov had chosen the exact polaron energy value, as it is done in ref. [ 25 ] for the case of strong coupling, or Feynman energy, as in ref. [ 26 ] for intermediate coupling, then even in the limit $\varepsilon_\infty/\varepsilon_0 = 0$, the two-center bipolaron energy would have been higher than double polaron energy. Work [54] differs from earlier work [60] only in the number of graphs illustrating the dependence of the intermediate coupling contribution on the crystal parameters. The contribution itself is calculated by V. K. Mukhomorov with the use of the erroneous expression.

Incorrect treatment of Buymistrov-Pekar method and erroneous results obtained in calculations of the polaron energy by this method attracted our attention to work [62] where the method is used to investigate a polaron. As a result of variation of the polaron functional (obtained using Buymistrov-Pekar canonical transformation and averaging of the initial Hamiltonian over phonon variables) with respect to the electron WF, V. K. Mukhomorov gets a simple differential equation describing a polaron. Apart from the term corresponding to the electron kinetic energy, the equation contains a term $V_0 \exp(-\gamma r^2)$ which, according to V. K. Mukhomorov, corresponds to the effective potential of interaction between an electron and phonons. Parameters of the potential are determined as a result of the parametrical fitting with the use of the value of the variational parameter $\beta$ in the simplest Gaussian function $\Psi_p(r) \sim \exp(-\beta r^2)$. The parameter $\beta$ is determined by the variation of the initial functional.

To our opinion, the problem of parametrical fitting is the most exacting. The author should have demonstrated that as a result of solution of a Schrödinger equation obtained in this way, one can obtain, at least approximately, the same polaron energy and the WF that were used in deriving the fitting potential. In other words, he should have proved that the method of deriving an approximate differential equation is self-consistent. In [62] no numerical values are given for the variational parameter $\beta$, nor for the parameters $V_0$ and $\gamma$ of the fitting potential. The author of [62] carefully investigates the Schrödinger equation and arrives at a conclusion that the first bound state appears for the parameter $\alpha = 2.8$.

As a numerical illustration we refer to our calculations of the polaron energy carried out by Buymistrov-Pekar method for the parameter $\alpha = 2.9$ which slightly exceeds the critical value found in [62]. For the simplest WF of a polaron chosen in the form $\psi_p = \exp(-\beta r^2)$, we obtained the following values of various contributions into the polaron energy $E_p = \overline{T}_e + \overline{V}_{sc} + \overline{V}_{ic}$ (where $\overline{T}_e = -\langle \Psi_p | \Delta | \Psi_p \rangle$, $V_{sc}$ is the term of the potential energy of the strong coupling functional, $\overline{V}_{ic}$ is an intermediate coupling addition in the total functional): $\sqrt{\beta} = 4.09322 \cdot 10^{-5}$, $\overline{T}_e = 5.26354 \cdot 10^{-9}$, $\overline{V}_{sc} = -1.33942 \cdot 10^{-4}$, $\overline{V}_{ic} = -2.89987$, $E_p = -2.90000$ (Feynman units of energy and length are used). Therefore, for $\alpha = 2.9$, the intermediate coupling Buymistrov-Pekar method yields the values completely coinciding with those found by the weak coupling method, i.e. $E_p \approx -\alpha\hbar\omega$. This peculiarity of Buymistrov-Pekar method is discussed in [47], eq. (11)-(14). This limiting case is obtained for the electron WF $\Psi = const = V^{-1/2}$, where V is the crystal volume. Indeed, for $\alpha < 6$, the functional of Buymistrov-Pekar method [47] is practically independent of the choice of the WF. For any WF, whether it be chosen in the form of a sum of exponents, or one exponent, or a sum of Gaussians, or one Gaussian, or a constant value independent of $r$, the polaron energy is $E_p = -\alpha\hbar\omega$ with good accuracy. In view of the fact that, according to eq. (11) of work [47], in the case under consideration the matrix element is $\langle\Psi|\cos(kr)|\Psi\rangle \cong 0$, differential equation (8), given in [62] takes a simple form $-\hbar^2/2m^* \Delta\Psi = \lambda\Psi$. Therefore, for $\alpha < 6$, the method suggested by V. K. Mukhomorov yields a differential equation independent of electron-phonon interaction. The latter results from the fact that V. K. Mukhomorov discarded in the functional all the terms which being averaged over angular variables were independent of the electron coordinate, however just these terms lead to a weak coupling formula for polaron functional.

In the limit of strong coupling, on the contrary, in equation (8) from [62], we can approximately put $a_k \cong 0$, then equation (8) studied by V.K. Mukhomorov in [62] goes over into Schrödinger equation for an electron moving in a field of self-consistent polarization potential:

$$\Delta\Psi_1 + \left(2\alpha_c \int (\Psi_2^2/r_{12}) d\tau_2 + \lambda\right)\Psi_1 = 0, \tag{4}$$

where $\Psi_i \equiv \Psi(r_i)$, and WF $\Psi_2$ is determined as a result of minimization of the polaron functional corresponding to the total energy of the studied system.

In the limit of strong coupling, the main equation (9) in [62] must have the same form. However, it does not, since the square bracket in the denominator under the integral sign lacks a square, a multiplier by the integral contains a superfluous factor $2\alpha_c^2$, and the unnecessary multiplier $\alpha_c^2$ appears at the term in the denominator of subintegral expression proportional to the wave vector.

In view of lots of errors in the basic equation (9) used in work [62], below we present a correct polaron equation which we obtained after varying expression (10) from work [47] with respect to the electronic WF and averaging it over angular variables.

$$\Delta\Psi + \left[\frac{4\alpha_c}{\pi}\int \frac{k^3 \sin(kr) J_k}{r(k^2+1-J_k^2)^2} dk + \lambda\right]\Psi = 0, \tag{5}$$

where $J_k = \langle\Psi(r)|\exp(i\mathbf{kr})|\Psi(r)\rangle$; for $\Psi \sim \exp(-\beta r^2)$ matrix element $J_k$ equals $\exp(-k^2/8\beta)$, $\alpha_c = e^2 c/2\hbar\omega a_p$ is the dimensionless constant of electron-phonon interaction, $c = \varepsilon_\infty^{-1} - \varepsilon_0^{-1}$, $a_p = \sqrt{(\hbar/2m\omega)}$ is the effective Feynman polaron radius.

The extreme case of the strong coupling can be obtained from (5), if in the denominator of subintegral expression we put $1 - J_k^2 \approx 0$, that corresponds to the approximate relationship

$$a_k = -\frac{C_k(1-J_k^2)}{\hbar^2 k^2/2m + \hbar\omega(1-J_k^2)} \approx 0, \tag{6}$$

where $C_k = -e\sqrt{4\pi\hbar\omega c}/k$, $c = 1/n^2 - 1/\varepsilon_0$, $n = \sqrt{\varepsilon_\infty}$ is the refractive index, or in the dimensionless Feynman units ($2m = 1$, $\hbar = 1$, $\hbar\omega = 1$) $C_k = \sqrt{8\pi\alpha_c}/k$.

Expression (6) is taken from work [47] (see formula (13) and (17)), because in work [62] commented by us, the coefficient $C_k$ and expression (6) determining a variation parameter are written down with errors. Thereby in the limit of the strong coupling equation (5) rearranges to the form

$$\Delta\Psi + \left[\frac{4\alpha_c}{\pi}\int \frac{\sin(kr) J_k}{rk} dk + \lambda\right]\Psi = 0. \tag{7}$$

If in equation (4) we substitute the expression $r_{12}^{-1} = \int \left(\exp(i\mathbf{k}\mathbf{r}_{12})/2\pi^2 k^2\right)$, integrate over the coordinate $\mathbf{r}_2$ and average over the angular variables of the wave vector, we will get, as one would expect, an expression which coincides with equation (7).

Therefore we believe that Fredholm integral equation studied in [62] bears no relation to Buymistrov-Pekar method and the polaron problem.

V. K. Mukhomorov states that the method of electron configurations interaction that he uses for Schrödinger states of electrons in a combined polarization potential well (which are obtained from one-electron eigenvalue equations) yields that nodeless state $1s$ occurs over the $2p$ - level on the energy scale. This contradiction allegedly disappears due to the fact that an additional restriction, namely virial theorem, is placed from below on the functional energy. We believe that this fact is caused by the same reason for which Buymistrov-Pekar method in the case of lack of translation symmetry [47] yields lower energies of the polaron energy than Feynman method does, which is reported in [54, 60, 62]. However, when one needs to expand the region of a bipolaron existence, the values of bipolaron energy calculated by V. K. Mukhomorov in [54, 60] by Buymistrov-Pekar method exceed not only those calculated by Feynman method, but also the values calculated by the perturbation theory technique. That is V. K. Mukhomorov makes numerous numerical errors which lead him to results inconsistent with reason.

## ABOUT ERRONEOUS TREATMENT OF LITERATURE ON BIPOLARON THEORY IN V. K. MUKHOMOROV'S WORKS

In [25] we did not state that the one-center bipolaron model is generally accepted, as V. K. Mukhomorov treats our words. Works [25-30], as well as [63], where integration over trajectories is used, just deal with the two-center bipolaron model. In these works the distance between the centers of polarization wells of the two polarons is considered as a variational parameter, but the energy minimum realizes for the one-center configuration rather than for the two-center one. In [25] we wrote that in view of a considerable energy gain obtained for the one-center bipolaron, after publication of [31] investigations of the two-center configuration practically ceased. The only exception is numerous works by V. K. Mukhomorov. The author of [33] states that the situation is directly opposite and in all the known works done by alternative methods a bipolaron in the ground singlet state is established to be a two-center axially-symmetrical dimer. By way of example he refers to papers [44, 64−69]. Let us analyze these works and show that the publications cited in [33] can serve to illustrate the validity of our statement, rather than prove V. K. Mukhomorov's contention. Among the papers cited in [33] only three, namely [64, 67, 68], deal with a two-center bipolaron.

In [64] (which was published much earlier than [31]) a model Hamiltonian is suggested for calculations of the bipolaron energy by integration over trajectories. There a variational parameter is introduced which can be treated as a distance between the centers of polarization wells of the two polarons. The model is considered qualitatively and numerical calculations are lacking.

In the well-known and frequently cited work [65] the bipolaron coupling energy is calculated by integration over trajectories in the framework of a one-center model, but not in the context of a two-center one, as it is stated in [33]. In the limit of strong coupling the bipolaron energy obtained in [65] coincides with the value found by the strong coupling method with the use of WF (1) for $a_1=a_3$, $a_2 \neq 0$, $R = 0$. It is emphasized ([65], p.252) that in the strong coupling limit the best results are obtained with the use of the WF presented in [31], and also with the WFs of the form of

$$\Psi(r_1, r_2) \sim (1 + k | r_1 - r_2 |) \exp(-\delta(r_1^2 + r_2^2)),$$
$$\Psi(r_1, r_2) \sim (1 - k \exp(-\varepsilon(r_1 - r_2)^2) \exp(-\delta(r_1^2 + r_2^2)),$$

which enable one to describe the correlation effects more correctly than WF (1) does. Attention is drawn to the fact that in WF (1) correlation effects are taken into account only if $a_2 \neq 0$, otherwise a bipolaron is not formed.

Later on [63], the results of [64, 65] were generalized by introducing an additional variational parameter $\langle a \rangle$ corresponding to the distance at which the interacting electrons fluctuate. This parameter is an analog of the distance between the centers of polarization wells of two polarons. Variational calculations revealed that the minimum of the functional corresponds to $\langle a \rangle = 0$. This result was obtained by the method of integration over trajectories. It completely correlates with our conclusion that a two-center bipolaron is unstable.

In [66] an indirect interaction of two electrons through a field of optical and acoustical phonons is investigated by the excitation theory method. The main conclusion of the work is that in ionic crystals an effective electron-electron interaction induces attraction of the electrons and screening of the Coulomb repulsion between the electrons. Therefore when an electron-phonon interaction is taken into account, the "inoculating" Coulomb repulsion $1/\varepsilon_\infty r$ decreases and becomes equal to $1/\varepsilon_0 r$. Against the background of the residual interaction $1/\varepsilon_0 r$ the potential of interaction between two electrons oscillates. These oscillations are caused by cutoff of the phonon spectrum by Debye value of the phonon wave vector.

It should be noted that in [66] the energy of interaction between two polarons is not calculated, since averaging is performed only over the coordinate corresponding to the center of mass of the interacting particles while the coordinate corresponding to the distance between the electrons remains free. Therefore in the limit of weak coupling the authors of [66] calculate a screened potential of interaction between two electrons and do not deal with finding the coupling energy of a one-center or two-center bipolaron.

In [67] the energy of a two-center bipolaron is found by a variational method. Exchange effects are taken into account by introducing an exchange-correlation pseudo potential used in atomic and molecular systems. The WF is chosen in the simplest form corresponding to Hartry-Fok approximation. Consideration of correlation effects leads to lowering of the two-center bipolaron energy. At $R = 0$ a nonzero coupling energy appears, however the one-center configuration corresponds to the maximum on the curve for the energy dependence on the distance R between the centers of polarization wells of two polarons.

Our calculations yield a similar dependence of the bipolaron energy on the distance R for a probe WF determined by a sum of several terms of the form of (1) for the parameters $a_{2i} = 0$, $a_{1i} \neq a_{3i}$ (i=1,2…N, N is the number of the terms). In [29] we give a dependence of the bipolaron energy on the distance R for the case of intermediate coupling ($\alpha = 9, \eta = 0$) corresponding to this approximation. When the interelectronic correlations ($a_{2i} \neq 0$, $a_{1i} \neq a_{3i}$) are taken into account still further, the extremum corresponding to the two-center configuration disappears. We believe that introduction of a more versatile pseudo potential which would take into account peculiarities of interaction between the electrons and a phonon field and specifics of the direct dependence of a probe WF on the distance between the electrons could improve considerably the results of [67] where the findings of [57, 58] are perfected only slightly.

In [68] a two-center bipolaron is studied by a simplest method of molecular orbitals. Configuration interaction is not taken into account. A later work of the same author [44] deals with a one-center bipolaron. The minimum found there is much deeper than that for a two-center configuration while in absolute magnitude it is nearly equal to that obtained in [31].

In paper [69] the properties of polaron gas are considered with due regard for interaction between polarons. The authors deal with weak and intermediate (from the side of weak $\alpha < 7$) electron-phonon coupling. In the limit of weak coupling the polaron gas is studied by the excitation theory method. At first, the coordinate of the bipolaron center of mass is excluded by the method suggested in [61], then the effective potential of pair interelectronic interaction is studied as a function of the distance $r_{12}$ between the electrons. Here, as in [66], averaging over the distance between the electrons is not carried out, therefore the object of investigations is not the total energy of the system consisting of two polarons, but a pair interelectronic potential shielded by phonons. Therefore the dependencies given in fig.1 of paper [69] remind (in the region of large distances between electrons) the dependence of the total energy of a two-center bipolaron on the distance between the centers of polarization wells only in shape. In passing on to systems with weak coupling, the curve changes into a hyperbole describing the dependence $1/r_{12}$.

Larsen [32] studies a screened interelectronic potential in the limit of weak electron-phonon coupling by Buymistrov-Pekar method. After averaging over the phonon variables, Adamovski's method of canonical transformation [70] also enables one to identify a pair (repulsive) interelectronic interaction corresponding to effective interaction between the electrons screened by a phonon field. While in [69] a pair interparticle potential is called an effective self-consistent potential of interelectronic correlation, in [70] this interaction is called interpolaron. However in the subsequent text it is explained that averaging of this interaction over the electron coordinates does not yield the total energy of the two polarons, since the main contribution into the bipolaron energy is made by the terms in the total functional which are independent of the distance between the electrons. These terms are responsible for the formation of a coupled bipolaron state.

Buymistrov-Pekar method (from the strong coupling region) and strong coupling methods do not enable one to identify this interaction since the coordinates of the center of mass cannot be separated from the distance between the electrons. In [71] these variables are separated, however only due to the fact that the author erroneously excludes the terms hindering this separation from the bipolaron functional, considering them to be fast oscillating. We believe that in carrying out variational calculations this approach is invalid and can be justified only when the missing terms cannot lead to an increase in the total energy. But if we bear in mind that in the strong coupling approximation the phonon spectrum is not restricted to Debye wave vector and integration over phonon variables and over electron coordinates is performed within infinite limits, then the origin of coordinates can be chosen at any point. In this case the operator of electron-phonon interaction can contain arbitrarily fast oscillating terms, which, according to the logic of the author of [71], can be neglected, including the operator itself. After "separation of the variables" V. K. Mukhomorov suggests a numerical solution of the equation obtained in this way, and as a zero approximation he uses the results of [54, 60] the fallacy of which we have already discussed.

E. P. Solodovnikova and A. N. Tavkhelidze [72] had generalized Bogolubov-Tyablikov translation invariant approach to a two-particle system long before paper [71] was published. At the end of their paper the authors express profound gratitude to N. N. Bogolubov for numerous and fruitful discussions of the problem of two bodies in the limit of strong coupling. Besides, well before the publication of [71] the bipolaron energy was calculated in the framework of a one-center model for coordinates of the center of mass and the distance between the electrons. The authors of [73] showed that the problem can be reduced to solution of two integro-differential equations. They explain in detail why one fails to separate the variables. In the iteration procedure used to find a numerical solution of the equations obtained, they use WFs, found in [43] by a variational method, as a zero approximation. The bipolaron energy obtained was lower by a few percent than that found earlier [43].

V. I. Vinetsky (an author of pioneering works which proved the possibility of the existence of a bound bipolaron [57, 58] in the framework of a two-center model) devoted his last work (co-authored with his followers) to calculations of the one-center bipolaron energy [74]. There he reproduced the results of [31] and calculated the bipolaron energy in terms of the center of mass and relative distance between the electrons. Use was made of both Gaussian and hydrogen-like WFs. One of the main conclusions of the work was that consideration of the correlation associated with a direct dependence of the WF of the system on the distance between the electrons considerably improves the criteria of a bipolaron existence.

The aim of our calculations [25-29] was not only to get qualitative estimates but also to obtain reliable numerical results. Thus, we applied our system of functions to calculate the energy of $H^-$ ion and the energy of para- and ortho- helium. The reliability of our approach was checked in calculations of the energy of a singlet and the nearest triplet terms of a hydrogen molecule. The results are given in [29] and represent one of the best numerical values of these quantities obtained thus far. The WFs that V. K. Mukhomorov applied to such classic systems of atomic and molecular physics yield only their qualitative estimates. This is especially true in regard to calculations of the energy of $H^-$ ion. In all the works aimed at getting exact energy values of the above mentioned quantities account is necessarily taken of the effects associated with the direct dependence of the WF of the system on the distance between the electrons. In paper [75] which provides an example of the most exact calculations of the energy of two-electron atomic systems, use is made of a system of WFs very close to ours, namely $\psi_{12} = \sum_{u} C_i (1 \pm P_{12}) \exp(-a_{1i} r_1 - 2a_{2i} r_{12} - a_{3i} r_i)$. This system is also used in papers [70, 76] the results of which are in a good agreement with the energies of a free bipolaron state [27-29] and a bound one [77] obtained by us by the intermediate coupling method.

V. K. Mukhomorov states that "if a one-center state of a continuum bipolaron were stable, additional consideration of the so-called ion terms in the electron configuration would lead to stabilization of the bipolaron". At the same time he refers to one of his works where simple calculations demonstrate that such corrections play no part in stabilization of bipolaron formations. We have reproduced the results of [31] and can state that calculation of the bipolaron energy with due regard for interelectronic correlations with the help of hydrogen-like functions used in [31] is an extremely cumbersome and labor consuming procedure. This is especially true in regard to analytical calculation of the integral corresponding to electron-phonon interaction. Therefore "simple calculations" presented by the author of [33] are erroneous. As for the fact that consideration of ion terms leads to stabilization of a bipolaron in a one-center configuration, this is just what we showed in [25].

We have not calculated the stability of biexcitons with regard for electronic correlations, therefore we cannot compare two-center and one-center models for this quantum-mechanical system. We can only give a reference to calculations of the biexciton energy carried out by Adamovski in the framework of a one-center model of an exciton and biexciton [78]. We believe that in view of a great difference between the effective masses of electrons and a hole, a biexciton may exist in a two-center configuration. The reason is that heavy holes repulse one another, so as protons in a hydrogen molecule do. This fact is reflected in the biexciton Hamiltonian. In a bipolaron the initial Hamiltonian does not have any repulsive terms, therefore the parallel that V. K. Mukhomorov draws between a biexciton and bipolaron is invalid.

Beneath criticism are V. K. Mukhomorov's remarks concerned with references to nonexistent experiments which allegedly confirm the preference of the two-center bipolaron model over the one-center one in metal ammonia solutions. Up to the present moment there is not any experimental evidence for the existence of a continuum polaron in these systems, not to speak of a two-center or one-center bipolaron. We can only speak of attempts to describe experimental facts in the framework of some or other model of which the polaron one, especially in continuum consideration, is not the main. This is indicated just in the monograph by Tompson which the author of [33] refers to.

We believe that after publication of paper [31], repeated reproduction and confirmation of its results by some independent research teams by various methods and with the use of various WFs chosen with regard for the direct dependence of the distance between the electrons, the author of [33], in turn, could have reproduced pertinent calculations and make sure that the virial relation, as it must be, holds true not only for the subsidiary minimum corresponding to the two-center configuration, but also for the one-center one, which leads to much lower energies of the bipolaron energy. Then he would not have done a series of erroneous works devoted to the study of vibrational and rotational spectra of "a continuum axial-symmetrical quasimolecular dimer" near the subsidiary minimum corresponding to the bipolaron two-center configuration in isotropic crystals. Numerous references to the works of this series are given in V. K. Mukhomorov's papers cited in this article.

## REFERENCES

1. Balabaev N.K., Lakhno V.D., Molchanov A.M., Atanasov B.P. *J. Mol. El.* 1990. **6** 155-166.
2. Chuev G.N., Lakhno V.D. *J. Theor. Biol.* 1993. **163**. 51-60.
3. Lakhno V.D., Chuev G.N. *Biofizika.* 1997. **41**. 313- 319.
4. Lakhno V.D., Chuev G.N *Himicheskaja fizika.* 1997. **16**. 50-54.
5. Lakhno V.D., Chuev G.N, Ustinin M.N. *Biofizika.* 1998. **43**. 999-952.
6. Lakhno V.D., Chuev G.N, Ustinin M.N.., Komarov V.M. *Biofizika.* 1998. **43**. 953-957.
7. Chuev G.N., Lakhno V.D., Ustinin M.N. *J. Biol. Phys.* 2000. **26**. 173-184.
8. Lakhno V.D. *Chemical Physics Letters.* 2007. **437**. 198-202.
9. Davydov A.S. *Uspehi Fizicheskih Nauk.* 1982. **138**. 603-643.
10. Davydov A.S. *Solitons in the molecular systems* . Kiev: Naukova dumka. 1984. 288 c.
11. Bogoljubov N. N.*Ukrainskij Matematicheskij Zhurnal..* 1950. **2**. 3-24.
12. Tjablikov S. V. *Zh. Éksp. Teor. Fiz.* 1951. **21**. 377-388.
13. Ivić Z, Zeković S, and Kostić D. *Phys. Rev. E.* 2002. **65**. 021911(1) - 021911(5).
14. Tsironis G.P., Molina M.I., and Hennig D. *Phys. Rev. E* 1994. **50**. 2365 – 2368.
15. Skott A. *Phys. Rep.* 1992. **217**. 1-68.
16. Lakhno V.D. *J. Biol. Phys.* 2000. **26**. 133-147.
17. Fialko N.S., Lakhno V.D. *Phys. Lett. A.* 2000. **278**. 108-111.
18. Fialko N.S., Lakhno V.D. *Regular & Chaotic Dynamics.* 2002. **7**. 299-313.
19. Lakhno V.D. Fialko N.S., *Pis'ma Zh. Éksp. Teor. Fiz.* 2003. **78**. 786-788.
20. Lakhno V.D. Fialko N.S. *Biofizika.* 2004. **49**. 575-578.
21. Korshunova A.N., Lakhno V.D. *Matem. modelirovaanie.* 2007. **19**. 3-13.
22. Lakhno V.D., Korshunova A.N. *Europ. Phys.J. B.* 2007. **55**. 85-87.
23. Iadonisi G., Mukhomorov V.K., Cantele G., Ninno D. *Phys. Rev. B.* 2005. **72**. 094305-1–094305-11.
24. Perroni C.A., Iadonisi G., Mukhomorov V.K. *Eur. Phys. J. B.* 2004. **41**. 163-170.
25. Kashirina N.I., Lakhno V.D., Sychyov V.V. *Fizika Tverdogo Tela.* 2003. **45**. 163-167.
26. Kashirina N.I., Lakhno V.D., Sychyov V.V. *Phys. stat. sol. (b).* 2003. **239.** 174-184.
27. Kashirina N.I., Lakhno V.D., Sychyov V.V. *Phys. stat. sol. (b).* 2002. **234.** 563-570.
28. Kashirina N.I., Lakhno V.D., Sychyov V.V. *Semiconductor Physics, Quantum Electronics & Optoelectronics.* 2002. **5.** 235-242.
29. Kashirina N.I., Lakhno V.D., Sychyov V.V. *Phys. Rev. B.* 2005. **71** (13). 134301-1–134301-13.
30. Kashirina N.I., Lakhno V.D. *Spatial configuration of the bipolyaron and the virial theorem. Fiz. Tverd. Tela.* 2008. **50**. 11-16.
31. Suprun S.G., Mojzhes B.JA. *Fiz. Tverd. Tel..* 1982. **24.** 1571-1573.
32. Larsen D.M. *Phys. Rev. B.* 1981. **23**. 628-631.
33. Mukhomorov V.K. *Fiz. Tverd. Tel.* 2006. **48.** 814- 820.
34. Bajmatov P.Zh., Huzhakulov D.Ch., Sharipov H.T. *Fiz. tverd. tel.* 1997. **39**. 284-285.
35. Gombash. P. P*roblem of many particles in quantum mechanics.* Moscow: Inostr. Lit. 1952. 280 c.
36. Löwdin P.O. *Adv. Chem. Phys.* 1959. **2.** 207-322.
37. Deigen M.F., Pekar S.I. *Zh. Éksp. Teor. Fiz.* 1951. **21.** 803-808.

38. Pekar S.I. *Issledovaniya po Elektronnoi Teorii Kristallov. (Research on the electronic theory of crystals)*. Moscow- Leningrad.: GITTL.1951. 256 c.
39. Pekar S.I. *Izbrannye trudy* (*Select labour*). Kiev: Naukova Dumka. 1988. 512 c.
40. Mitra T.K. *Phys. Let. A*. 1989. **142**. 398-400.
41. Adamowski J., Bednarek S. *J. Phys. Cond. Matter*. 1992. **4**. 2845-2855.
42. Dzhumanov S., Baratov A.A, Abboudy S. *Phys. Rev. B*. 1996. **54**. 13121-13128.
43. Verbist G., Smondyrev M.A., Peeters F.M., Devreese J.T. *Phys. Rev. B* 1992. **45**. 5262-5269.
44. Sahoo S. *J. Phys.: Cond. Matt.* 1995. **7.** 4457-4466.
45. Tomasevich O.F. *Zh. Éksp. Teor. Fiz.* 1951. **21**. 1223-1226.
46. Davydov A.S. Nauchnye zapiski Kievskogo gos. universiteta. **XI** (IV}. Tr. fiz. fak. 1952. **6.** 5-10.
47. Buĭmistrov V. M. and Pekar S. I., *Zh. Éksp. Teor. Fiz.* 1957. **32**, 1193-1199. [Sov. Phys. JETP 1957. **5**. 970-976]
48. Kashirina N.I., Mozdor E.V., Pashickii É.A., Sheka V.I. *Izv. Rossijskoj Akademii nauk. Serija Fizicheskaja.* 1995. **59.** 127-133.
49. Mott N.F. *J. Phys. C*. 1993. **5**. 3487-3506.
50. Slater J. C. *Quantum Theory of Molecules and Solids*. Vol. 1: *Electronic Structure of Molecules*. New York: McGraw-Hill. 1963. [Russian translation] Moscow: Mir. 1965. 588 p.
51. Bethe H. A. *Quantum Mechanics of the Simplest Systems. [Russian translation]*. Leningrad- Moscow: ONTI. 1935. 400 p.
52. Mukhomorov V.K. *Optika i Spektr.* 1979. **46**. 926 -929.
53. Gabdoulline R.R. *Dokl. Ross. Akad. Nauk.* 1993. **333**. 23-27.
54. Mukhomorov V.K. *J. Phys. C.* **13.** 2001. 3633-3642.
55. Mukhomorov V.K. *Optika i spektr.*1993. **74**. 1083-1104.
56. Mukhomorov V.K. *Phys. stat. sol. (b).* 2002. **231**. 462-476.
57. Vineckii V.L., Gitterman M.Sh. *Zh. Éksp. Teor. Fiz.* 1957. **33**. 730-734.
58. Vineckii V.L. *Zh. Éksp. Teor. Fiz.*1961. **40**. 1459-1468.
59. Deigen M.F. *Zh. Éksp. Teor. Fiz.* 1951. **21.** 992-1000.
60. Mukhomorov V.K. *Optika i spektr.* 1999. **86.** 50-55.
61. Lee T.D., Low F.E., Pines D. *Phys. Rev.* 1953. **90**. 297-302.
62. Mukhomorov V.K. *Fiz. tverd. tela*. 2000. **42**. 1559-1562.
63. Verbist G., Peeters F.M., and Devreese J.T. *Phys. Rev. B* 1991. **43**. 2712-2720.
64. Massumi T. *Suppl. Progr. Teor. Phys.* 1975. **57.** 22-34.
65. Hiramoto H., Toyozawa Y. *J. Phys. Soc. Japan.* 1985. **54**. 245-259.
66. Bishop M.F., Overhauser A.W. *Phys. Rev.B* 1981. **23**. 3627-3637.
67. Glushkov A.I. *Zh. Fizich. Him.* 1990. **64.** 1579-1581.
68. Sahoo S. *Phys. Lett. A.* 1994. **195**. 105-109.
69. De Fillips G., Cataudella V., Iadonisi G. *Europ. Phys. J. B* 1999. **8**. 339-351.
70. Adamowski J. *Phys. Rev. B* 1989. **39**. 3649-3652.
71. Mukhomorov V.K. *Fiz. Tverd. Tela*. 2002. **44**. 232-238.
72. Solodovnikova E.P., Tavhelidze A.N. *Teor. i Mat. Fiz.* 1974. **21**. 13-24.
73. Qinghu C., Kelin W., Shaolong W. *Phys. Rev. B* 1994. **50**. 164-167.
74. Vineckii V.L., Meredov O., Janchuk V.Ja. *Teor. Eksp. Him.* 1989. **25.** 641-647.
75. Thakkar A.J., Smit V.H. *Jr. Phys. Rev. A*. 1977. **15.** 1-15.
76. Adamowski J. *Phys. Rev. B* 1989. **39**. 13061-13066.
77. Kashirina N.I., Lakhno V.D., Sychev V.V. *Fiz. Teh. Poluprov.* 2003. **37**. 318-322.
78. Adamowski J., Bednarec S., Suffczynski M. *Sol. St. Com.* 1978. **25.** 89-92.